\documentclass[twocolumn]{aastex701}
\usepackage{amsmath}
\usepackage{amssymb}	
\usepackage{CJK}
\usepackage{float}
\usepackage{booktabs}
\usepackage{color}
\usepackage{multirow}
\usepackage{graphicx}
\usepackage{subcaption}

\begin{document}
\begin{CJK*}{UTF8}{gbsn}
\title{On the Formation of GW231123 in Population III Star Clusters}

\correspondingauthor{Long Wang}
\email{wanglong8@sysu.edu.cn}

\author[orcid=0000-0002-1197-2054,sname='Liu']{Shuai Liu (刘帅)}
\affiliation{School of Electronic and Electrical Engineering, Zhaoqing University, Zhaoqing 526061, People's Republic of China}
\email[]{}  

\author[orcid=0000-0001-8713-0366,sname='Wang']{Long Wang (王龙)} 
\affiliation{School of Physics and Astronomy, Sun Yat-sen University, Daxue Road, Zhuhai, 519082, People's Republic of China}
\affiliation{CSST Science Center for the Guangdong-Hong Kong-Macau Greater Bay Area, Zhuhai, 519082, People's Republic of China}
\email[]{wanglong8@sysu.edu.cn}

\author[orcid=0000-0002-8461-5517,sname=Tanikawa]{Ataru Tanikawa}
\affiliation{Center for Information Science, Fukui Prefectural University, 4-1-1 Matsuoka Kenjojima, Eiheiji-cho, Fukui 910-1195, Japan}
\email{}

\author{Weiwei Wu (吴维为)}
\affiliation{School of Physics and Astronomy, Sun Yat-sen University, Daxue Road, Zhuhai, 519082, People's Republic of China}
\email{}

\author[orcid=0000-0002-6465-2978, sname=Fujii]{Michiko S. Fujii}
\affiliation{Department of Astronomy, Graduate School of Science, The University of Tokyo, 7-3-1 Hongo, Bunkyo-ku, Tokyo 113-0033, Japan}
\email{}




\begin{abstract}
GW231123 is a binary black hole merger whose primary component lies within or above the pair-instability mass gap, while the secondary component falls within this gap. The standard theory of stellar evolution is significantly challenged by this event. We investigate the formation of candidate progenitors of GW231123 in Population III (Pop III) star clusters. We find that they could form through stellar mergers, binary black hole mergers, and mixed mergers. The mass distribution of these candidate progenitors covers the component masses of GW231123. Under our model assumptions, their predicted merger rate density spans the range of $0.001-0.26{\rm Gpc^{-3}yr^{-1}}$, encompassing that of GW231123. These findings suggest that GW231123 may originate from Pop III star clusters. Furthermore, such candidate progenitors are expected to be detectable by future gravitational wave detectors LISA/Taiji/TianQin/DECIGO/Cosmic Explorer/Einstein Telescope, which would provide valuable insights into the formation scenarios of events like GW231123.

\end{abstract}




\section{Introduction} \label{sec:introduction}
The LIGO-Virgo-KAGRA Collaboration released the heaviest binary black hole (BBH) merger GW231123 to date \citep{LIGOScientific:2025rsn}. This event is consistent with the coalescence of two BHs with component masses of $m_{1}=137_{-17}^{+22}M_{\odot}$ and $m_{2}=103_{-52}^{+20}M_{\odot}$, occurring at redshift of $z=0.39_{-0.24}^{+0.27}$. The primary mass $m_{1}$ lies within or exceeds the pair-instability (PI) mass gap of $\sim60-130M_{\odot}$ \citep{Farmer:2019jed,Farmer:2020xne,Woosley:2021xba,Hendriks:2023yrw}, while the secondary mass $m_{2}$ falls within this gap, presenting a significant challenge to standard stellar evolution theory. The resulting merger remnant is an intermediate-mass BH of $225_{-43}^{+26}M_{\odot}$, along with the merger rate density of $0.08_{-0.07}^{+0.19}{\rm Gpc^{-3}yr^{-1}}$. However, the formation of such a high-mass system as GW231123 remains an open question.

Multiple formation scenarios are proposed to explain the presence of BHs within the PI mass gap (PIBHs). One possibility involves isolated, metal-poor Pop III stars, which may retain their hydrogen envelopes due to weak stellar winds and directly collapse into PIBHs \citep{Mapelli:2019ipt,Spera:2018wnw}. In close Pop III binaries with rapid rotation and short orbital periods, processes such as chemically homogeneous evolution could prevent merger and enable the formation of PIBHs \citep{Kinugawa:2014zha,Mandel:2015qlu,Marchant:2016wow,Kinugawa:2020xws,Tanikawa:2024mpj,Popa:2025dpz}. Population I/II stars undergoing runaway collisions in dense environments, such as young star clusters (YSCs) \citep{Mapelli:2016vca,DiCarlo:2019fcq,Kremer:2020wtp,Gonzalez:2020xah,Rizzuto:2021atw,Costa:2022aka} or globular star clusters (GCs) \citep{ArcaSedda:2023mlv}, or compact multiple systems \citep{Renzo:2020smh}, may form very massive stars (VMSs) that evolve into PIBHs. Alternatively, PIBHs may result from hierarchical BH mergers in dynamical environments including YSCs \citep{Mapelli:2020xeq,Sedda:2021abh}, GCs \citep{Miller:2001ez,Rodriguez:2019huv,Li:2025pyo,Paiella:2025qld}, and nuclear star clusters \citep{Antonini:2018auk,Fragione:2020nib,Kritos:2022non,Mahapatra:2024qsy,Delfavero:2025lup,Li:2025fnf,Stegmann:2025cja}, or the accretion disks of active galactic nuclei \citep{Bartos:2016dgn,Stone:2016wzz,Mckernan:2017ssq,Yang:2019cbr,Tagawa:2019osr,McKernan:2019beu,Vaccaro:2023cwr,Sedda:2023big}. Accretion processes in star clusters \citep{Bartos:2025pkv,Kiroglu:2025vqy,Roupas:2025djb} and direct core collapse \citep{Croon:2025gol,Gottlieb:2025ugy} may also contribute to the formation of PIBHs. Another intriguing possibility is that these objects are primordial black holes formed in the early universe \citep{Bird:2016dcv,Bird:2022wvk,Clesse:2016vqa,Clesse:2020ghq,DeLuca:2025fln,Yuan:2025avq}.
     
Pop III star clusters are also proposed as a promising formation environment for PIBHs \citep{Wang:2022unj,Liu:2023zea,2025ApJ...986..163W}. In such clusters, VMS could form through multiple stellar collisions and subsequently evolve into PIBHs. BBHs may also form via binary stellar evolution and merge to produce PIBHs. Alternatively, BBHs assembled through dynamical processes, such as gravitational capture and exchange interactions, could also result in PIBHs. If Pop III clusters are embedded within mini dark matter halos, these halos could shield the clusters from galactic tidal disruption, allowing BBHs containing PIBHs to persist and merge from high redshifts to the present Universe. In the companion paper \citep{Tanikawa:2025fxw}, we also examine the possibility that isolated Pop III binary stars form BBH mergers with PIBHs like GW231123.

In this work, we first investigate the possibility that GW231123 originated from Pop III star clusters. In Sec. \ref{sec:method}, we describe the models adopted for Pop III star clusters. The results are presented in Sec. \ref{sec:result}, and our conclusions are summarized in Sec. \ref{sec:conclusion}.

\section{Method}\label{sec:method}
Our analysis is based on data from $N$-body simulations of Pop III star clusters embedded in mini dark matter halos \citep{Liu:2023zea,2025ApJ...986..163W}. The initial conditions of the clusters follow the model `NFWden\_long\_w9\_noms\_imf1' described in \citep{Wang:2022unj}, which tends to produce more merging BBHs. Six models of the clusters are constructed by combining different cluster masses and primordial binary fractions (PBFs), which are summarized in Table \ref{tab:model}. For each model, the metallicities of stars are set to be $Z=2\times10^{-10}$. The initial masses of stars follow a top-heavy initial mass function (IMF) with a power-law index of -1, ranging from $1M_{\odot}$ to $150M_{\odot}$, following the studies from \citep{2016MNRAS.462.1307S,Chon:2021jlx,2024MNRAS.530.2453C} and \citep{Latif:2021xad}. The initial period, mass ratios and eccentricities of primordial binaries follow Sana's distributions \citep{Sana:2012px}. The initial mean values of escape velocities from clusters and mini dark matter halos, and half-mass densities across different models are listed in Table \ref{tab:model}. The potential of mini dark matter halos is assumed to follow the model in \citep{Navarro:1995iw}, where the virial mass and radius are $4\times10^{7}M_{\odot}$ and 280pc. The remaining initial parameters are described in \citep{Liu:2023zea,2025ApJ...986..163W}. Single and binary stars are evolved using the fast population synthesis codes \textsc{seemp} and \textsc{bseemp} \citep{Tanikawa:2019isi,Tanikawa:2021qqi}. Note that, for the single star evolution model in \textsc{bseemp}, we choose the L model \citep{Tanikawa:2021qqi}, which is impossible to form a BBH merger with PIBHs through isolated binary evolution. Thus, all the BBH mergers with PIBHs are formed through dynamical interactions. The dynamical evolution of the clusters over 12\,Gyr is simulated with the $N$-body code \textsc{petar} \citep{2020MNRAS.497..536W}. The influence of galactic potential on the clusters is implemented by \textsc{petar} using \textsc{galpy} \citep{Bovy:2014vfa}.

Among the data from $N$-body simulations for Pop III star clusters, we define the merging BBHs with a final merger mass of $225_{-43}^{+26}M_{\odot}$ as candidate progenitors of GW231123, according to the 90\% probability intervals for these parameters inferred from GW observation.

\begin{table*}
    \centering
    \caption{The first row lists the models of Pop III star clusters. The second, third and fourth rows show the cluster mass $M_{\rm clu}$, PBF and the number of simulations $N_{\rm sim}$, respectively. The fifth and sixth rows present the initial mean values of escape velocities from clusters and mini dark matter halos, and half-mass densities, with units of km/s and $M_{\odot}{\rm pc}^{-3}$. The seventh and eighth rows present the total number $N_{\rm total}$ and mean number $N_{\rm mean}$ of candidate progenitors of GW231123 across the simulations. The fractions of candidate progenitors formed through different channels (stellar mergers, BBH mergers, and mixed mergers) are listed in the ninth, tenth, eleventh rows, respectively. Explanations of the formation channels are provided in Fig. \ref{fig:mass distribution}. The twelfth row shows the merger rate densities for candidate progenitors averaged over redshift. The last two rows give the fraction of candidate progenitors that could not form due to kick velocities of merger remnants, and the resulting merger rate densities in units of ${\rm Gpc}^{-3}{\rm yr}^{-1}$, respectively. The symbol ``$-$" in the sixth column indicates that no candidate progenitors of GW231123 formed in the M1000-PBF0 model.}
    \label{tab:model}
    \hspace{-0cm} 
    \begin{tabular}{ccccccc}
        \toprule
        Model & M100000-PBF0 & M100000-PBF1 & M10000-PBF0 & M10000-PBF1 & M1000-PBF0 & M1000-PBF1 \\
        \midrule
        $M_{\rm clu} (M_{\odot})$ & 100000 & 100000 & 10000 & 10000 & 1000 & 1000 \\
        PBF & 0 & 1 & 0 & 1 & 0 & 1 \\
        $N_{\rm sim}$ & 168 & 168 & 275 & 275  & 275 & 275 \\
        $\bar{v}_{\rm esc}$ & 103.1 & 103.1 & 68.5 & 68.4 & 57.8 & 58.1 \\
        $\bar{\rho}_{\rm h}$ & 11934.2 & 11910.6 & 1242.6 & 1210.0 & 218.9 & 381.9\\
        $N_{\rm total}$  & 27 & 98 & 8 & 42 & -- & 1 \\
        $N_{\rm mean}$  & 0.16 & 0.58 & 0.03 & 0.15 & -- & 0.004 \\
        $f_{\rm stellar\ merger}$ & 0.22 & 0.79 & 1.0 & 0.67 & -- & 1.0\\
        $f_{\rm BBH\ merger}$ & 0.11 & 0.10 & 0.0 & 0.02 & -- & 0.0\\
        $f_{\rm mixed\ merger}$ & 0.67 & 0.11 & 0.0 & 0.31 & -- & 0.0\\
        $\mathcal{R}_{\rm lower}-\mathcal{R}_{\rm upper}$ & 0.004-0.03 & 0.02-0.1 & 0.008-0.05 & 0.04-0.3 & -- & 0.01-0.07\\
        $f_{\rm kick}$ & 0.67 & 0.08 & 0.0 & 0.14 & -- & 0.0 \\
        $(\mathcal{R}_{\rm lower}-\mathcal{R}_{\rm upper})_{\rm kick}$ & 0.001-0.01 & 0.018-0.09 & 0.008-0.05 & 0.034-0.26 & -- & 0.01-0.07\\
        \bottomrule
    \end{tabular}
\end{table*}

\section{Result}\label{sec:result}

The average numbers of candidate progenitors of GW231123 across different Pop III models are listed in Table \ref{tab:model}. When PBF=0, about 0.16 candidate progenitors are formed per cluster with a mass of $100000M_{\odot}$. As the cluster mass decreases, the average number of candidate progenitors steadily declines, and no candidate progenitors are found in clusters with a mass of $1000M_{\odot}$. This trend arises because more massive clusters contain more stars and have higher stellar densities, making the formation of candidate progenitors more likely. When PBF=1, the average number of candidate progenitors increases significantly, since the presence of primordial binaries enhances the probability of their formation.

\begin{figure}[htbp]
  \centering
  \begin{subfigure}{0.95\linewidth}
    \centering
    \includegraphics[width=\linewidth,keepaspectratio]{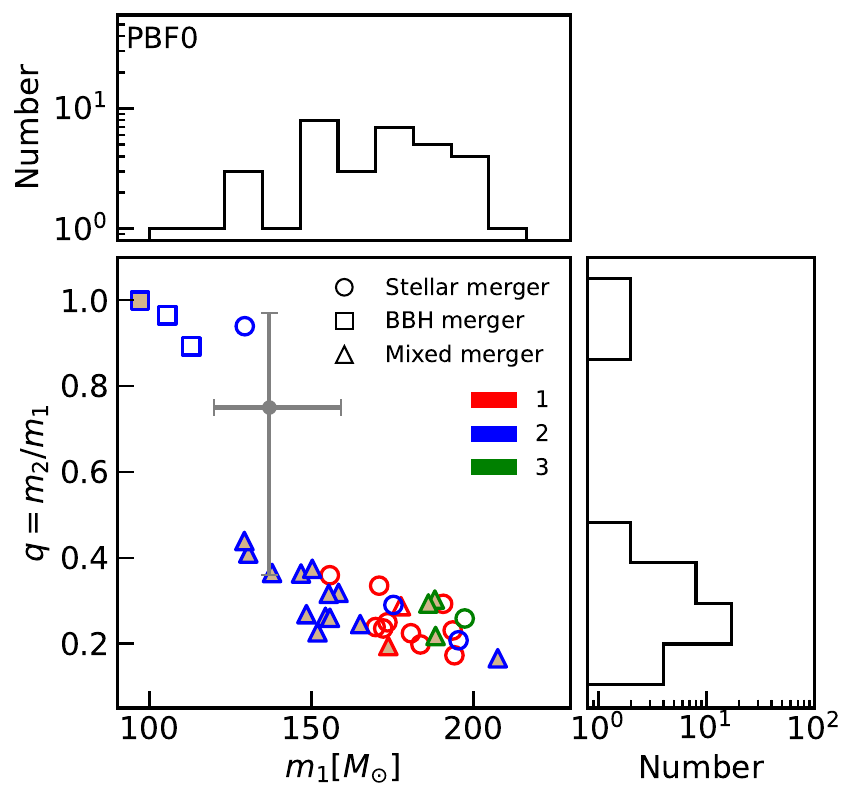}
  \end{subfigure}
  \vspace{0.6em}
  \begin{subfigure}{0.95\linewidth}
    \centering
    \includegraphics[width=\linewidth,keepaspectratio]{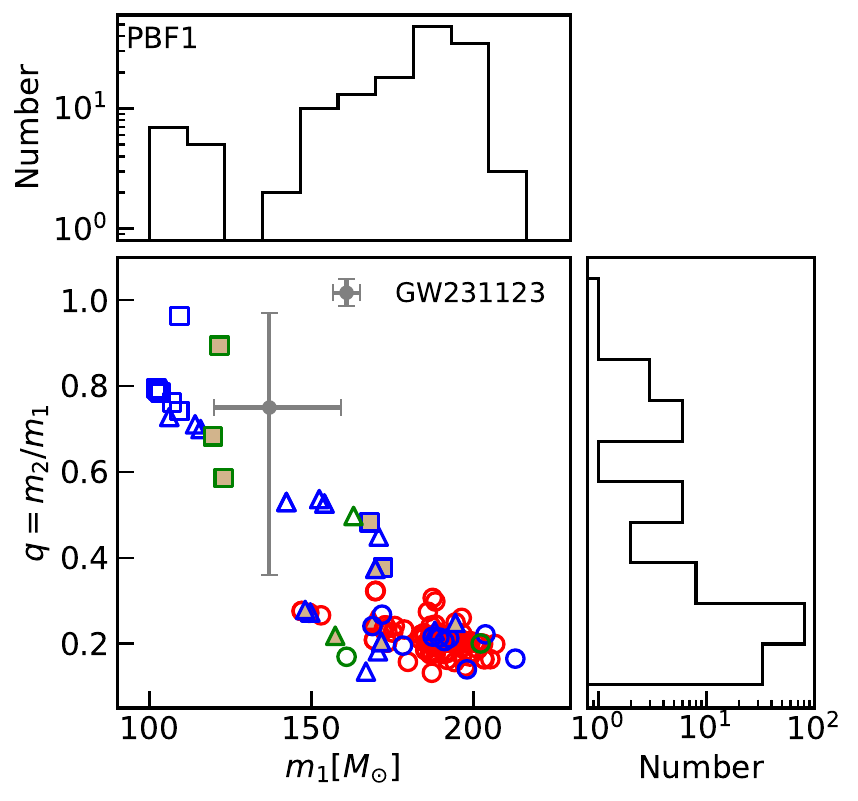}
  \end{subfigure}
  \caption{Distributions of the primary mass $m_{1}$ and mass ratio $q=m_{2}/m_{1}$ for candidate progenitors of GW231123. The upper and lower panels show results from models of Pop III clusters with PBF=0 and 1, respectively. Symbols indicate the formation channels of candidate progenitors. Circles and squares correspond to cases where both of $m_{1}$ and $m_{2}$ originate from stellar mergers (the massive components are PIBHs evolved from massive stars that underwent stellar mergers, thereby avoiding pair-instability supernovae) and BBH mergers, respectively. Triangles represent mixed mergers, i.e., $m_{1}$ and $m_{2}$ are from stellar mergers and BBH mergers, respectively. Symbols with a tan facecolor denote candidate progenitors whose components are ejected from their host clusters due to kick velocities exceeding the clusters’ escape velocities. Colors indicate the total number of mergers contributing to the formation of candidate progenitors, defined as the sum of the mergers producing $m_{1}$ and those producing $m_{2}$ ($n_{1}+n_{2}$), with red, blue and green denoting 1, 2 and 3, respectively. Note that almost $m_{2}$ originates from the evolution of ordinary single stars in stellar merger and mixed merger channels, i.e., $n_{2}=0$. The gray dot marks the median value of GW231123, with the error bar indicating the 90\% probability intervals inferred from GW observations.}
  \label{fig:mass distribution}
\end{figure}

The mass distributions of candidate progenitors of GW231123 are shown in Fig. \ref{fig:mass distribution}. For both cases of ${\rm PBF}=0$ and 1, the primary mass $m_{1}$ spans the range of $100-200M_{\odot}$, with a concentration between $150M_{\odot}$ and $200M_{\odot}$. The mass ratio $q=m_{2}/m_{1}$ ranges from 0.1--1, with a concentration around 0.2. These mass distributions encompass the parameters of GW231123, which has $m_{1}=137_{-17}^{+22}M_{\odot}$ and $q=0.75_{-0.39}^{+0.22}$.

The formation channels of candidate progenitors of GW231123 are indicated in Fig. \ref{fig:mass distribution}, and their fractions in each model are listed in Table \ref{tab:model}. Our results indicate that the candidate progenitors could form through 1-3 repeated mergers, including stellar mergers, BBH mergers and mixed mergers ($m_{1}$ and $m_{2}$ from stellar mergers and BBH mergers respectively), with the first and third channels contributing to the largest fractions in both cases of PBF=0 and 1. The massive components of candidate progenitors of GW231123 formed through the stellar merger channel are PIBHs originating from massive stars with very large hydrogen envelopes and He-core masses below $45M_{\odot}$ that underwent stellar mergers, which allowed them to avoid pair-instability supernovae. The fraction of stellar mergers tends to increase in low-mass clusters. This is because stars, having larger radii than BHs, possess much larger collision cross-sections, making stellar mergers more likely to occur than BBH mergers in low-mass clusters with lower number densities. Furthermore, since stellar mergers do not receive gravitational wave (GW) kick velocities, their contribution is independent of the escape velocity of host clusters. The candidate progenitors of GW231123 formed through stellar mergers and mixed mergers tend to have large $m_{1}$ and small $q$. This is because the massive components are from stellar mergers, whereas $m_{2}$ evolves from ordinary single stars or formed through single BBH mergers, resulting in $m_{1}\gg m_{2}$.

The BBH merger formation channel of candidate progenitors of GW231123 can be understood as follows. A top-heavy IMF in Pop III clusters results in a higher fraction of massive stars and, consequently, more BH progenitors. BHs formed from the evolution of massive stars in Pop III clusters are heavier due to extremely low or zero metallicities. As a result, the candidate progenitors of GW231123 could form through only two BBH mergers, fewer than typically required in metal-rich clusters \citep{Antonini:2018auk,Fragione:2020nib,Antonelli:2023gpu}. This would decrease the likelihood that GW recoil kicks eject BBH mergers before they could continue merging with BHs to become candidate progenitors of GW231123. The candidate progenitors from BBH mergers occupy the region with smaller $m_{1}$ and larger $q$, since both $m_{1}$ and $m_{2}$ are remnants of single BBH mergers and therefore have comparable but generally lower masses.

During the formation of $m_{1}$ and $m_{2}$ through BBH mergers, the merger remnants will receive kick velocities, which depend on the spins and mass ratios of the BBHs. If these kick velocities exceed the escape velocities of the host clusters, the merger remnants will be ejected. Note that \textsc{petar} does not track BH spins, although \textsc{bseemp} evolves the spins of stars and their remnant BHs. \citet{2025ApJ...986..163W} estimate the spins of BBHs in Pop III star clusters using a post-processing method. They reference the BH spins from isolated binaries evolved with \textsc{bseemp}. Using the same method, we estimate the kick velocities of BBH merger remnants during the formation of candidate progenitors of GW231123. We also calculate the escape velocities of Pop III clusters, accounting for both stars and the mini dark matter halos, which can reach $\sim50{\rm km/s}-\sim100{\rm km/s}$, as listed in Table \ref{tab:model}. Although some candidate progenitors of GW231123 formed through BBH mergers are ejected from clusters, as shown in Fig. 1 and listed in Table \ref{tab:model}, a significant fraction remains bound to the clusters. After removing these expelled candidate progenitors, the remaining mass distributions still encompass the parameter space of GW231123.

\begin{figure}[h]
\centering
    \includegraphics[width=0.45\textwidth]{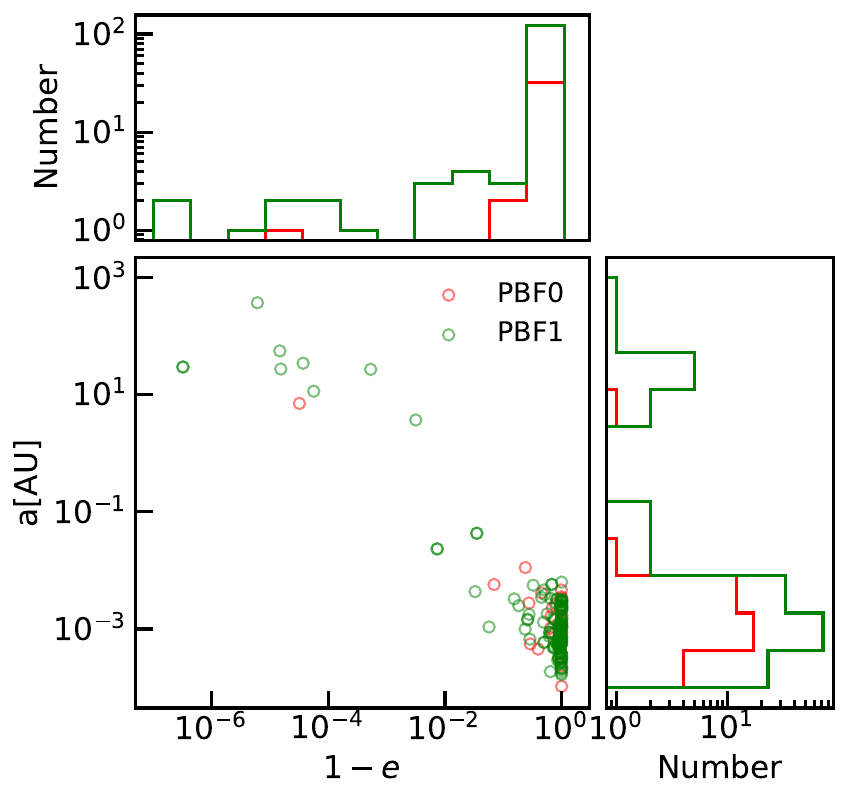}
	\caption{The orbital element (eccentricity $e$ and semi-major axis $a$) distributions of candidate progenitors of GW231123 at their formation time, defined as the moment BBHs emerge during binary star evolution. The red and green dots represent the results from different models with PBF=0 and 1, respectively.}
\label{fig:e semi distribution}
\end{figure}

The eccentricity $e$ and semi-major axis $a$ distributions of candidate progenitors of GW231123 at their formation time are represented in Fig. \ref{fig:e semi distribution}. In both cases with PBF=0 and 1, most of the systems have nearly circular orbits, with $e\sim0$, although a small fraction exhibit high eccentricities, with $1-e<10^{-2}$. The semi-major axes span a wide range of $10^{-4}-10^{3}{\rm AU}$, indicating that the BBHs are initially far from the merger phase. These results suggest that candidate progenitors of GW231123 undergo significant orbital circularization due to GW emission between formation and merger. Consequently, their eccentricities are expected to be close to zero in the LIGO/Virgo/KAGRA frequency bands, consistent with the assumption that GW231123 originated from a circular BBHs merger, as adopted in the analysis of its GW signal.

\begin{figure}[h]
\centering
    \includegraphics[width=0.45\textwidth]{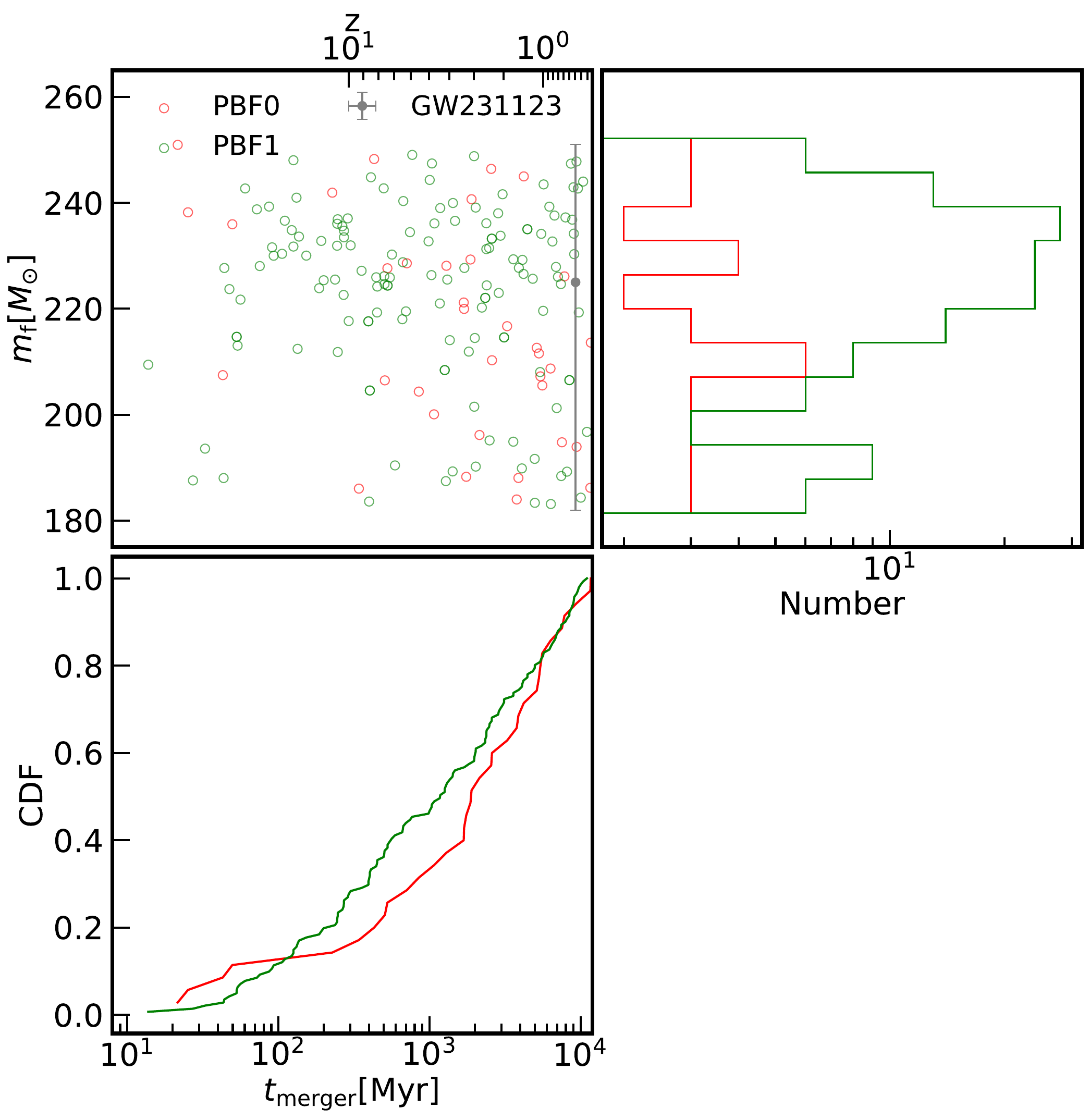}
	\caption{Merger remnant masses $m_{\rm f}$ and merger time $t_{\rm merger}$ of candidate progenitors of GW231123. The merger time $t_{\rm merger}$ refers to the moment during the evolution of Pop III star clusters, at which the mergers occur. Red and green dots represent the results from different models with PBF=0 and 1, respectively.}
\label{fig:mf cdf tmerger}
\end{figure}

The merger time $t_{\rm merger}$ of candidate progenitors of GW231123 during the Pop III star cluster evolution, along with their merger remnant masses $m_{\rm f}$ are shown in Fig. \ref{fig:mf cdf tmerger}. In both of situations with ${\rm PBF}=0$ and 1, $m_{\rm f}$ are concentrated in the range of 210--240$M_{\odot}$, covering the median final masses of GW231123, which is 225\,$M_{\odot}$. Most of the candidate progenitors of GW231123 mergers ($\sim80\%$) occur at redshift $z>2$. If the mini dark matter halos that shield Pop III star clusters from disruption by the galactic potential survive to the present day, a fraction of candidate progenitors of GW231123 could merge as late as the present epoch, covering the redshift $z=0.39_{-0.24}^{+0.27}$ of GW231123. The upper and lower limits of merger rate density $\mathcal{R}$ averaged over redshift, estimated by Eq. (7) in \citep{Liu:2023zea}, are also provided in Table \ref{tab:model}. The values of $\mathcal{R}$ could range from $0.004\,{\rm Gpc^{-3}yr^{-1}}$ to $0.3\,{\rm Gpc^{-3}yr^{-1}}$ across different models, encompassing the GW231123 merger rate density of $0.08_{-0.07}^{+0.19}\,{\rm Gpc^{-3}yr^{-1}}$ inferred by GW observations. Taking into account the impact of kick velocities of BBH merger remnants into account, the resulting merger rate densities of candidate progenitors, listed in Table \ref{tab:model}, range from $0.001\,{\rm Gpc^{-3}yr^{-1}}$ to $0.26\,{\rm Gpc^{-3}yr^{-1}}$, which is still consistent with that inferred for GW231123.

\begin{figure}[h]
\centering
    \includegraphics[width=0.48\textwidth]{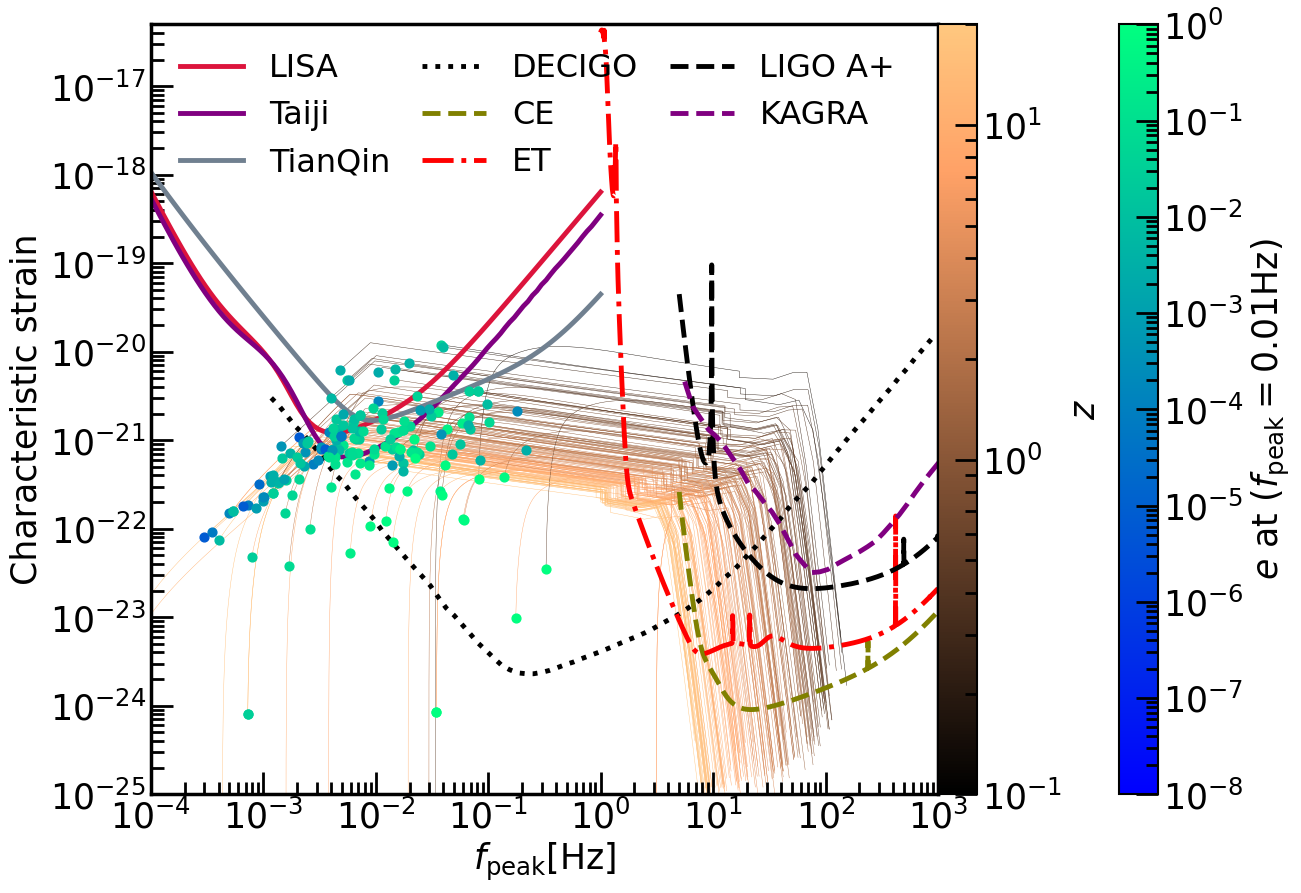}
	\caption{Peak frequency $f_{\rm peak}$ and the corresponding characteristic strain $h_{\rm cn_{peak}}$ of GWs emitted by candidate progenitors of GW231123 in all the models. The copper and winter color bars represent redshift $z$ and eccentricities at peak $f_{\rm peak}=0.01$\,Hz, respectively. Sensitivity curves of GW detectors LISA, Taiji, TianQin, DECIGO, CE, ET, LIGO, and KAGRA are represented by different colors and line styles.}
\label{fig:hcn fpeak}
\end{figure}

Following the calculation on GWs in \citep{Liu:2023zea}, we compute the peak frequencies $f_{\rm peak}$ and the corresponding characteristic strains $h_{cn_{\rm peak}}$ of GWs emitted by candidate progenitors of GW231123, as shown in Fig. \ref{fig:hcn fpeak}. A significant fraction of these sources is potentially detectable by future mHz space-borne GW detectors, LISA \citep{2017arXiv170200786A}, Taiji \citep{Ruan:2018tsw}, TianQin \citep{Liu:2020eko}, as their $h_{cn_{\rm peak}}$ values exceed the sensitive curves of the detectors at certain frequency bands. These detectable sources are typically located at low redshift ($z<2$), or exhibit low eccentricities ($e<0.01$) at $f_{\rm peak}=0.01$Hz, where LISA/Taiji/TianQin are most sensitive. Candidate progenitors of GW231123 located at low redshift would also be detectable by ground-based GW detectors like LIGO. Furthermore, nearly all such progenitors are covered by future detectors DECIGO \citep{Kawamura:2011zz}, Cosmic Explorer (CE) \citep{Reitze:2019iox}, and Einstein Telescope (ET) \citep{Punturo:2010zz}. Furthermore, we find that several events have high eccentricities at $f_{\rm peak}=0.01$\,Hz. They are detectable by LIGO/Virgo/KAGRA/CE/ET/DECIGO, while they are not by LISA/Taiji/TianQin. In contrast, isolated Pop III binary stars, which typically evolve into circular orbits, cannot form such events. In other words, if GW231123 originated from isolated Pop III binary stars, then similar events discovered in the future would be detected by both LIGO/Virgo/KAGRA/CE/ET/DECIGO and LISA/Taiji/TianQin. This observational feature may be a valuable clue to identify the origin of GW231123.

\section{Conclusion}\label{sec:conclusion}
In this work, we first investigate the possibility that GW231123 originated from Pop III star clusters, using the $N$-body simulation data from \citep{Liu:2023zea} and \citep{2025ApJ...986..163W}. We show that Pop III star clusters efficiently produce GW231123-like events. The candidate progenitors of GW231123 are produced through 1-3 repeated mergers, which can be classified into three distinct channels: stellar mergers, mixed mergers, and BBH mergers. Among these, the first two channels contribute the largest fractions, with the fraction of stellar mergers tending to increase in low-mass clusters. Since stellar mergers are not subject to ejection by GW recoil kicks, their contribution remains unaffected. The top-heavy IMF leads to a higher fraction of massive stars capable of evolving into BHs. In metal-poor environments, BHs formed from massive stars are generally more massive. As a result, the candidate progenitors of GW231123 could form through only two BBH mergers, fewer than the number typically required in metal-rich clusters. Although BBH merger remnants receive GW recoil kicks, the combined escape velocities of clusters and their host mini dark matter halos, reaching $\sim50{\rm km/s}-\sim100{\rm km/s}$, allow a significant fraction of the candidate progenitors of GW231123 to be retained within clusters.

The primary mass $m_{1}$ and mass ratio $q$ of candidate progenitors of GW231123 generally lie in the ranges of $100$--$200M_{\odot}$ and $0.1$--$1$, respectively, encompassing the observed values of GW231123. Most candidate progenitors are expected to be circular in LIGO/Virgo/KAGRA bands, having undergone orbital circularization due to GW emission from formation to merger. This is consistent with the assumption adopted in the GW231123 signal analysis. About 80\% of candidate progenitor mergers occur at redshifts of $z>2$, while the rest extend to as late as the present epoch, if the mini dark matter halos that protect clusters from disruption by galactic potential survive to the present day, covering the redshift of GW231123. Most components of candidate progenitors of GW231123 are expected to form via BBH mergers and possess high spins. The merger rate density averaged over redshift of candidate progenitors of GW231123, accounting for the kicks from spin binary mergers, ranges from $0.001{\rm Gpc^{-3}yr^{-1}}$ to $0.26{\rm Gpc^{-3}yr^{-1}}$, which is consistent with the inferred value of GW231123. Candidate progenitors of GW231123 located at low redshifts ($z<2$) would be detectable by LIGO/Virgo/KAGRA. Among them, those with low eccentricities ($e<0.01$) in the mHz band would also be observable by LISA/Taiji/TianQin, while those with high eccentricities would not. Therefore, by performing multiband observations of sources with high eccentricities, it may be possible to determine whether GW231123-like events originate from Pop III star clusters or from isolated Pop III binary stars. Furthermore, almost candidate progenitors would be covered by DECIGO/CE/ET. 

Since the existence of Pop III star clusters has not yet been confirmed observationally, although our analysis suggests that GW231123 may have originated from Pop III star clusters, we cannot definitively conclude that this is the case. Alternative formation scenarios and environments may also contribute to the GW231123 population. With the future GW detectors LISA/Taiji/TianQin/DECIGO/CE/ET, we would detect candidate progenitors of GW231123 with high redshift or during their early inspiral phases. These observations would provide crucial insights into the origin and evolutionary history of the GW231123 population.   

\begin{acknowledgments}

S.L. thanks the support from Natural Science Foundation of China (grant No. 12503054), Zhaoqing City Science and Technology Innovation Guidance Project (No. 241216104168995), and the Young Faculty Research Funding Project of Zhaoqing University (No. qn202518). L.W. thanks the support from the National Natural Science Foundation of China through grant 21BAA00619, 12573041 and 12233013, the High-level Youth Talent Project (Provincial Financial Allocation) through the grant 2023HYSPT0706, and the one-hundred-talent project of Sun Yat-sen University, and the Fundamental Research Funds for the Central Universities, Sun Yat-sen University (2025QNPY04). A.T. thanks the Grants-in-Aid for Scientific Research (17H06360, 24K07040 and 25K01035), and the Step-up program at Fukui Prefectural University.

\end{acknowledgments}


\bibliography{ref}{}

\begin{thebibliography}{}
\expandafter\ifx\csname natexlab\endcsname\relax\def\natexlab#1{#1}\fi
\providecommand{\url}[1]{\href{#1}{#1}}
\providecommand{\dodoi}[1]{doi:~\href{http://doi.org/#1}{\nolinkurl{#1}}}
\providecommand{\doeprint}[1]{\href{http://ascl.net/#1}{\nolinkurl{http://ascl.net/#1}}}
\providecommand{\doarXiv}[1]{\href{https://arxiv.org/abs/#1}{\nolinkurl{https://arxiv.org/abs/#1}}}

\bibitem[{LIG(2025)}]{LIGOScientific:2025rsn}
 2025, \bibinfo{title}{{GW231123: a Binary Black Hole Merger with Total Mass
  190-265 $M_{\odot}$},} \doarXiv{2507.08219}

\bibitem[{P. {Amaro-Seoane} {et~al.}(2017){Amaro-Seoane}, {Audley}, {Babak},
  {Baker}, {Barausse}, {Bender}, {Berti}, {Binetruy}, {Born}, {Bortoluzzi},
  {Camp}, {Caprini}, {Cardoso}, {Colpi}, {Conklin}, {Cornish}, {Cutler},
  {Danzmann}, {Dolesi}, {Ferraioli}, {Ferroni}, {Fitzsimons}, {Gair}, {Gesa
  Bote}, {Giardini}, {Gibert}, {Grimani}, {Halloin}, {Heinzel}, {Hertog},
  {Hewitson}, {Holley-Bockelmann}, {Hollington}, {Hueller}, {Inchauspe},
  {Jetzer}, {Karnesis}, {Killow}, {Klein}, {Klipstein}, {Korsakova}, {Larson},
  {Livas}, {Lloro}, {Man}, {Mance}, {Martino}, {Mateos}, {McKenzie},
  {McWilliams}, {Miller}, {Mueller}, {Nardini}, {Nelemans}, {Nofrarias},
  {Petiteau}, {Pivato}, {Plagnol}, {Porter}, {Reiche}, {Robertson},
  {Robertson}, {Rossi}, {Russano}, {Schutz}, {Sesana}, {Shoemaker}, {Slutsky},
  {Sopuerta}, {Sumner}, {Tamanini}, {Thorpe}, {Troebs}, {Vallisneri},
  {Vecchio}, {Vetrugno}, {Vitale}, {Volonteri}, {Wanner}, {Ward}, {Wass},
  {Weber}, {Ziemer}, \& {Zweifel}}]{2017arXiv170200786A}
{Amaro-Seoane}, P., {Audley}, H., {Babak}, S., {et~al.} 2017,
  \bibinfo{title}{{Laser Interferometer Space Antenna},} arXiv e-prints,
  arXiv:1702.00786, \dodoi{10.48550/arXiv.1702.00786}

\bibitem[{A. Antonelli {et~al.}(2023)Antonelli, Kritos, Ng, Cotesta, \&
  Berti}]{Antonelli:2023gpu}
Antonelli, A., Kritos, K., Ng, K. K.~Y., Cotesta, R., \& Berti, E. 2023,
  \bibinfo{title}{{Classifying the generation and formation channels of
  individual LIGO-Virgo-KAGRA observations from dynamically formed binaries},}
  Phys. Rev. D, 108, 084044, \dodoi{10.1103/PhysRevD.108.084044}

\bibitem[{F. Antonini {et~al.}(2019)Antonini, Gieles, \&
  Gualandris}]{Antonini:2018auk}
Antonini, F., Gieles, M., \& Gualandris, A. 2019, \bibinfo{title}{{Black hole
  growth through hierarchical black hole mergers in dense star clusters:
  implications for gravitational wave detections},} Mon. Not. Roy. Astron.
  Soc., 486, 5008, \dodoi{10.1093/mnras/stz1149}

\bibitem[{M. Arca~Sedda {et~al.}(2023)Arca~Sedda, Kamlah, Spurzem, Rizzuto,
  Naab, Giersz, \& Berczik}]{ArcaSedda:2023mlv}
Arca~Sedda, M., Kamlah, A. W.~H., Spurzem, R., {et~al.} 2023,
  \bibinfo{title}{{The dragon-II simulations {\textendash} II. Formation
  mechanisms, mass, and spin of intermediate-mass black holes in star clusters
  with up to 1 million stars},} Mon. Not. Roy. Astron. Soc., 526, 429,
  \dodoi{10.1093/mnras/stad2292}

\bibitem[{I. Bartos \& Z. Haiman(2025)Bartos \& Haiman}]{Bartos:2025pkv}
Bartos, I., \& Haiman, Z. 2025, \bibinfo{title}{{Accretion is All You Need:
  Black Hole Spin Alignment in Merger GW231123 Indicates Accretion Pathway},}
  \doarXiv{2508.08558}

\bibitem[{I. Bartos {et~al.}(2017)Bartos, Kocsis, Haiman, \&
  M{\'a}rka}]{Bartos:2016dgn}
Bartos, I., Kocsis, B., Haiman, Z., \& M{\'a}rka, S. 2017,
  \bibinfo{title}{{Rapid and Bright Stellar-mass Binary Black Hole Mergers in
  Active Galactic Nuclei},} Astrophys. J., 835, 165,
  \dodoi{10.3847/1538-4357/835/2/165}

\bibitem[{S. Bird {et~al.}(2016)Bird, Cholis, Mu{\~n}oz, Ali-Ha{\"\i}moud,
  Kamionkowski, Kovetz, Raccanelli, \& Riess}]{Bird:2016dcv}
Bird, S., Cholis, I., Mu{\~n}oz, J.~B., {et~al.} 2016, \bibinfo{title}{{Did
  LIGO detect dark matter?},} Phys. Rev. Lett., 116, 201301,
  \dodoi{10.1103/PhysRevLett.116.201301}

\bibitem[{S. Bird {et~al.}(2023)Bird {et~al.}}]{Bird:2022wvk}
Bird, S., {et~al.} 2023, \bibinfo{title}{{Snowmass2021 Cosmic Frontier White
  Paper: Primordial black hole dark matter},} Phys. Dark Univ., 41, 101231,
  \dodoi{10.1016/j.dark.2023.101231}

\bibitem[{J. Bovy(2015)Bovy}]{Bovy:2014vfa}
Bovy, J. 2015, \bibinfo{title}{{galpy: A Python Library for Galactic
  Dynamics},} Astrophys. J. Suppl., 216, 29, \dodoi{10.1088/0067-0049/216/2/29}

\bibitem[{S. {Chon} {et~al.}(2024){Chon}, {Hosokawa}, {Omukai}, \&
  {Schneider}}]{2024MNRAS.530.2453C}
{Chon}, S., {Hosokawa}, T., {Omukai}, K., \& {Schneider}, R. 2024,
  \bibinfo{title}{{Impact of radiative feedback on the initial mass function of
  metal-poor stars},} \mnras, 530, 2453, \dodoi{10.1093/mnras/stae1027}

\bibitem[{S. Chon {et~al.}(2021)Chon, Omukai, \& Schneider}]{Chon:2021jlx}
Chon, S., Omukai, K., \& Schneider, R. 2021, \bibinfo{title}{{Transition of the
  initial mass function in the metal-poor environments},} Mon. Not. Roy.
  Astron. Soc., 508, 4175, \dodoi{10.1093/mnras/stab2497}

\bibitem[{S. Clesse \& J. Garc{\'\i}a-Bellido(2017)Clesse \&
  Garc{\'\i}a-Bellido}]{Clesse:2016vqa}
Clesse, S., \& Garc{\'\i}a-Bellido, J. 2017, \bibinfo{title}{{The clustering of
  massive Primordial Black Holes as Dark Matter: measuring their mass
  distribution with Advanced LIGO},} Phys. Dark Univ., 15, 142,
  \dodoi{10.1016/j.dark.2016.10.002}

\bibitem[{S. Clesse \& J. Garcia-Bellido(2022)Clesse \&
  Garcia-Bellido}]{Clesse:2020ghq}
Clesse, S., \& Garcia-Bellido, J. 2022, \bibinfo{title}{{GW190425, GW190521 and
  GW190814: Three candidate mergers of primordial black holes from the QCD
  epoch},} Phys. Dark Univ., 38, 101111, \dodoi{10.1016/j.dark.2022.101111}

\bibitem[{G. Costa {et~al.}(2022)Costa, Ballone, Mapelli, \&
  Bressan}]{Costa:2022aka}
Costa, G., Ballone, A., Mapelli, M., \& Bressan, A. 2022,
  \bibinfo{title}{{Formation of black holes in the pair-instability mass gap:
  Evolution of a post-collision star},} Mon. Not. Roy. Astron. Soc., 516, 1072,
  \dodoi{10.1093/mnras/stac2222}

\bibitem[{D. Croon {et~al.}(2025)Croon, Sakstein, \& Gerosa}]{Croon:2025gol}
Croon, D., Sakstein, J., \& Gerosa, D. 2025, \bibinfo{title}{{Can stellar
  physics explain GW231123?},} \doarXiv{2508.10088}

\bibitem[{V. De~Luca {et~al.}(2025)De~Luca, Franciolini, \&
  Riotto}]{DeLuca:2025fln}
De~Luca, V., Franciolini, G., \& Riotto, A. 2025, \bibinfo{title}{{GW231123: a
  Possible Primordial Black Hole Origin},} \doarXiv{2508.09965}

\bibitem[{V. Delfavero {et~al.}(2025)Delfavero, Ray, Cook, Nathaniel, McKernan,
  Ford, Postiglione, McPike, \& O'Shaughnessy}]{Delfavero:2025lup}
Delfavero, V., Ray, S., Cook, H.~E., {et~al.} 2025, \bibinfo{title}{{Prospects
  for the formation of GW231123 from the AGN channel},} \doarXiv{2508.13412}

\bibitem[{U.~N. Di~Carlo {et~al.}(2020)Di~Carlo, Mapelli, Bouffanais, Giacobbo,
  Santoliquido, Bressan, Spera, \& Haardt}]{DiCarlo:2019fcq}
Di~Carlo, U.~N., Mapelli, M., Bouffanais, Y., {et~al.} 2020,
  \bibinfo{title}{{Binary black holes in the pair-instability mass gap},} Mon.
  Not. Roy. Astron. Soc., 497, 1043, \dodoi{10.1093/mnras/staa1997}

\bibitem[{R. Farmer {et~al.}(2020)Farmer, Renzo, de~Mink, Fishbach, \&
  Justham}]{Farmer:2020xne}
Farmer, R., Renzo, M., de~Mink, S., Fishbach, M., \& Justham, S. 2020,
  \bibinfo{title}{{Constraints from gravitational wave detections of binary
  black hole mergers on the
  $^{12}\rm{C}\left(\alpha,\gamma\right)^{16}\!\rm{O}$ rate},} Astrophys. J.
  Lett., 902, L36, \dodoi{10.3847/2041-8213/abbadd}

\bibitem[{R. Farmer {et~al.}(2019)Farmer, Renzo, de~Mink, Marchant, \&
  Justham}]{Farmer:2019jed}
Farmer, R., Renzo, M., de~Mink, S.~E., Marchant, P., \& Justham, S. 2019,
  \bibinfo{title}{{Mind the gap: The location of the lower edge of the pair
  instability supernovae black hole mass gap},}
  \dodoi{10.3847/1538-4357/ab518b}

\bibitem[{G. Fragione \& J. Silk(2020)Fragione \& Silk}]{Fragione:2020nib}
Fragione, G., \& Silk, J. 2020, \bibinfo{title}{{Repeated mergers and ejection
  of black holes within nuclear star clusters},} Mon. Not. Roy. Astron. Soc.,
  498, 4591, \dodoi{10.1093/mnras/staa2629}

\bibitem[{E. Gonz{\'a}lez {et~al.}(2021)Gonz{\'a}lez, Kremer, Chatterjee,
  Fragione, Rodriguez, Weatherford, Ye, \& Rasio}]{Gonzalez:2020xah}
Gonz{\'a}lez, E., Kremer, K., Chatterjee, S., {et~al.} 2021,
  \bibinfo{title}{{Intermediate-mass Black Holes from High Massive-star Binary
  Fractions in Young Star Clusters},} Astrophys. J. Lett., 908, L29,
  \dodoi{10.3847/2041-8213/abdf5b}

\bibitem[{O. Gottlieb {et~al.}(2025)Gottlieb, Metzger, Issa, Li, Renzo, \&
  Isi}]{Gottlieb:2025ugy}
Gottlieb, O., Metzger, B.~D., Issa, D., {et~al.} 2025,
  \bibinfo{title}{{Spinning into the Gap: Direct-Horizon Collapse as the Origin
  of GW231123 from End-to-End GRMHD Simulations},} \doarXiv{2508.15887}

\bibitem[{D.~D. Hendriks {et~al.}(2023)Hendriks, van Son, Renzo, Izzard, \&
  Farmer}]{Hendriks:2023yrw}
Hendriks, D.~D., van Son, L. A.~C., Renzo, M., Izzard, R.~G., \& Farmer, R.
  2023, \bibinfo{title}{{Pulsational pair-instability supernovae in
  gravitational-wave and electromagnetic transients},} Mon. Not. Roy. Astron.
  Soc., 526, 4130, \dodoi{10.1093/mnras/stad2857}

\bibitem[{S. Kawamura {et~al.}(2011)Kawamura {et~al.}}]{Kawamura:2011zz}
Kawamura, S., {et~al.} 2011, \bibinfo{title}{{The Japanese space gravitational
  wave antenna: DECIGO},} Class. Quant. Grav., 28, 094011,
  \dodoi{10.1088/0264-9381/28/9/094011}

\bibitem[{T. Kinugawa {et~al.}(2014)Kinugawa, Inayoshi, Hotokezaka, Nakauchi,
  \& Nakamura}]{Kinugawa:2014zha}
Kinugawa, T., Inayoshi, K., Hotokezaka, K., Nakauchi, D., \& Nakamura, T. 2014,
  \bibinfo{title}{{Possible Indirect Confirmation of the Existence of Pop III
  Massive Stars by Gravitational Wave},} Mon. Not. Roy. Astron. Soc., 442,
  2963, \dodoi{10.1093/mnras/stu1022}

\bibitem[{T. Kinugawa {et~al.}(2021)Kinugawa, Nakamura, \&
  Nakano}]{Kinugawa:2020xws}
Kinugawa, T., Nakamura, T., \& Nakano, H. 2021, \bibinfo{title}{{Formation of
  binary black holes similar to GW190521 with a total mass of $\sim
  150\,M_{\odot}$ from Population III binary star evolution},} Mon. Not. Roy.
  Astron. Soc., 501, L49, \dodoi{10.1093/mnrasl/slaa191}

\bibitem[{F. K{\i}ro{\u{g}}lu {et~al.}(2025)K{\i}ro{\u{g}}lu, Kremer, \&
  Rasio}]{Kiroglu:2025vqy}
K{\i}ro{\u{g}}lu, F., Kremer, K., \& Rasio, F.~A. 2025, \bibinfo{title}{{Beyond
  Hierarchical Mergers: Accretion-Driven Origins of Massive, Highly Spinning
  Black Holes in Dense Star Clusters},} \doarXiv{2509.05415}

\bibitem[{K. Kremer {et~al.}(2020)Kremer, Spera, Becker, Chatterjee, Di~Carlo,
  Fragione, Rodriguez, Ye, \& Rasio}]{Kremer:2020wtp}
Kremer, K., Spera, M., Becker, D., {et~al.} 2020, \bibinfo{title}{{Populating
  the upper black hole mass gap through stellar collisions in young star
  clusters},} Astrophys. J., 903, 45, \dodoi{10.3847/1538-4357/abb945}

\bibitem[{K. Kritos {et~al.}(2023)Kritos, Berti, \& Silk}]{Kritos:2022non}
Kritos, K., Berti, E., \& Silk, J. 2023, \bibinfo{title}{{Massive black hole
  assembly in nuclear star clusters},} Phys. Rev. D, 108, 083012,
  \dodoi{10.1103/PhysRevD.108.083012}

\bibitem[{M.~A. Latif {et~al.}(2022)Latif, Whalen, \& Khochfar}]{Latif:2021xad}
Latif, M.~A., Whalen, D., \& Khochfar, S. 2022, \bibinfo{title}{{The Birth Mass
  Function of Population III Stars},} Astrophys. J., 925, 28,
  \dodoi{10.3847/1538-4357/ac3916}

\bibitem[{G.-P. Li \& X.-L. Fan(2025)Li \& Fan}]{Li:2025pyo}
Li, G.-P., \& Fan, X.-L. 2025, \bibinfo{title}{{The Hierarchical Merger
  Scenario for GW231123},} \doarXiv{2509.08298}

\bibitem[{Y.-J. Li {et~al.}(2025)Li, Tang, Xue, \& Fan}]{Li:2025fnf}
Li, Y.-J., Tang, S.-P., Xue, L.-Q., \& Fan, Y.-Z. 2025,
  \bibinfo{title}{{GW231123: a product of successive mergers from $\sim 10 $
  stellar-mass black holes},} \doarXiv{2507.17551}

\bibitem[{S. Liu {et~al.}(2020)Liu, Hu, Zhang, \& Mei}]{Liu:2020eko}
Liu, S., Hu, Y.-M., Zhang, J.-d., \& Mei, J. 2020, \bibinfo{title}{{Science
  with the TianQin observatory: Preliminary results on stellar-mass binary
  black holes},} Phys. Rev. D, 101, 103027, \dodoi{10.1103/PhysRevD.101.103027}

\bibitem[{S. Liu {et~al.}(2024)Liu, Wang, Hu, Tanikawa, \& Trani}]{Liu:2023zea}
Liu, S., Wang, L., Hu, Y.-M., Tanikawa, A., \& Trani, A.~A. 2024,
  \bibinfo{title}{{Merging hierarchical triple black hole systems with
  intermediate-mass black holes in population III star clusters},} Mon. Not.
  Roy. Astron. Soc., 533, 2262, \dodoi{10.1093/mnras/stae1946}

\bibitem[{P. Mahapatra {et~al.}(2024)Mahapatra, Chattopadhyay, Gupta, Antonini,
  Favata, Sathyaprakash, \& Arun}]{Mahapatra:2024qsy}
Mahapatra, P., Chattopadhyay, D., Gupta, A., {et~al.} 2024,
  \bibinfo{title}{{Reconstructing the Genealogy of LIGO-Virgo Black Holes},}
  Astrophys. J., 975, 117, \dodoi{10.3847/1538-4357/ad781b}

\bibitem[{I. Mandel \& S.~E. de~Mink(2016)Mandel \& de~Mink}]{Mandel:2015qlu}
Mandel, I., \& de~Mink, S.~E. 2016, \bibinfo{title}{{Merging binary black holes
  formed through chemically homogeneous evolution in short-period stellar
  binaries},} Mon. Not. Roy. Astron. Soc., 458, 2634,
  \dodoi{10.1093/mnras/stw379}

\bibitem[{M. Mapelli(2016)Mapelli}]{Mapelli:2016vca}
Mapelli, M. 2016, \bibinfo{title}{{Massive black hole binaries from runaway
  collisions: the impact of metallicity},} Mon. Not. Roy. Astron. Soc., 459,
  3432, \dodoi{10.1093/mnras/stw869}

\bibitem[{M. Mapelli {et~al.}(2021)Mapelli, Santoliquido, Bouffanais, Sedda,
  Artale, \& Ballone}]{Mapelli:2020xeq}
Mapelli, M., Santoliquido, F., Bouffanais, Y., {et~al.} 2021,
  \bibinfo{title}{{Mass and Rate of Hierarchical Black Hole Mergers in Young,
  Globular and Nuclear Star Clusters},} Symmetry, 13, 1678,
  \dodoi{10.3390/sym13091678}

\bibitem[{M. Mapelli {et~al.}(2020)Mapelli, Spera, Montanari, Limongi, Chieffi,
  Giacobbo, Bressan, \& Bouffanais}]{Mapelli:2019ipt}
Mapelli, M., Spera, M., Montanari, E., {et~al.} 2020, \bibinfo{title}{{Impact
  of the Rotation and Compactness of Progenitors on the Mass of Black Holes},}
  Astrophys. J., 888, 76, \dodoi{10.3847/1538-4357/ab584d}

\bibitem[{P. Marchant {et~al.}(2016)Marchant, Langer, Podsiadlowski, Tauris, \&
  Moriya}]{Marchant:2016wow}
Marchant, P., Langer, N., Podsiadlowski, P., Tauris, T.~M., \& Moriya, T.~J.
  2016, \bibinfo{title}{{A new route towards merging massive black holes},}
  Astron. Astrophys., 588, A50, \dodoi{10.1051/0004-6361/201628133}

\bibitem[{B. McKernan {et~al.}(2020)McKernan, Ford, O'Shaughnessy, \&
  Wysocki}]{McKernan:2019beu}
McKernan, B., Ford, K. E.~S., O'Shaughnessy, R., \& Wysocki, D. 2020,
  \bibinfo{title}{{Monte Carlo simulations of black hole mergers in AGN discs:
  Low $\chi_{\rm eff}$ mergers and predictions for LIGO},} Mon. Not. Roy.
  Astron. Soc., 494, 1203, \dodoi{10.1093/mnras/staa740}

\bibitem[{B. Mckernan {et~al.}(2018)Mckernan {et~al.}}]{Mckernan:2017ssq}
Mckernan, B., {et~al.} 2018, \bibinfo{title}{{Constraining Stellar-mass Black
  Hole Mergers in AGN Disks Detectable with LIGO},} Astrophys. J., 866, 66,
  \dodoi{10.3847/1538-4357/aadae5}

\bibitem[{M.~C. Miller \& D.~P. Hamilton(2002)Miller \&
  Hamilton}]{Miller:2001ez}
Miller, M.~C., \& Hamilton, D.~P. 2002, \bibinfo{title}{{Production of
  intermediate-mass black holes in globular clusters},} Mon. Not. Roy. Astron.
  Soc., 330, 232, \dodoi{10.1046/j.1365-8711.2002.05112.x}

\bibitem[{J.~F. Navarro {et~al.}(1996)Navarro, Frenk, \&
  White}]{Navarro:1995iw}
Navarro, J.~F., Frenk, C.~S., \& White, S. D.~M. 1996, \bibinfo{title}{{The
  Structure of cold dark matter halos},} Astrophys. J., 462, 563,
  \dodoi{10.1086/177173}

\bibitem[{L. Paiella {et~al.}(2025)Paiella, Ugolini, Spera, Branchesi, \&
  Sedda}]{Paiella:2025qld}
Paiella, L., Ugolini, C., Spera, M., Branchesi, M., \& Sedda, M.~A. 2025,
  \bibinfo{title}{{Assembling GW231123 in star clusters through the combination
  of stellar binary evolution and hierarchical mergers},} \doarXiv{2509.10609}

\bibitem[{S.~A. Popa \& S.~E. de~Mink(2025)Popa \& de~Mink}]{Popa:2025dpz}
Popa, S.~A., \& de~Mink, S.~E. 2025, \bibinfo{title}{{Very Massive, Rapidly
  Spinning Binary Black Hole Progenitors through Chemically Homogeneous
  Evolution -- The Case of GW231123},} \doarXiv{2509.00154}

\bibitem[{M. Punturo {et~al.}(2010)Punturo {et~al.}}]{Punturo:2010zz}
Punturo, M., {et~al.} 2010, \bibinfo{title}{{The Einstein Telescope: A
  third-generation gravitational wave observatory},} Class. Quant. Grav., 27,
  194002, \dodoi{10.1088/0264-9381/27/19/194002}

\bibitem[{D. Reitze {et~al.}(2019)Reitze {et~al.}}]{Reitze:2019iox}
Reitze, D., {et~al.} 2019, \bibinfo{title}{{Cosmic Explorer: The U.S.
  Contribution to Gravitational-Wave Astronomy beyond LIGO},} Bull. Am. Astron.
  Soc., 51, 035.
\newblock \doarXiv{1907.04833}

\bibitem[{M. Renzo {et~al.}(2020)Renzo, Cantiello, Metzger, \&
  Jiang}]{Renzo:2020smh}
Renzo, M., Cantiello, M., Metzger, B.~D., \& Jiang, Y.~F. 2020,
  \bibinfo{title}{{The Stellar Merger Scenario for Black Holes in the
  Pair-instability Gap},} Astrophys. J. Lett., 904, L13,
  \dodoi{10.3847/2041-8213/abc6a6}

\bibitem[{F.~P. Rizzuto {et~al.}(2022)Rizzuto, Naab, Spurzem, Arca-Sedda,
  Giersz, Ostriker, \& Banerjee}]{Rizzuto:2021atw}
Rizzuto, F.~P., Naab, T., Spurzem, R., {et~al.} 2022, \bibinfo{title}{{Black
  hole mergers in compact star clusters and massive black hole formation beyond
  the mass gap},} Mon. Not. Roy. Astron. Soc., 512, 884,
  \dodoi{10.1093/mnras/stac231}

\bibitem[{C.~L. Rodriguez {et~al.}(2019)Rodriguez, Zevin, Amaro-Seoane,
  Chatterjee, Kremer, Rasio, \& Ye}]{Rodriguez:2019huv}
Rodriguez, C.~L., Zevin, M., Amaro-Seoane, P., {et~al.} 2019,
  \bibinfo{title}{{Black holes: The next generation{\textemdash}repeated
  mergers in dense star clusters and their gravitational-wave properties},}
  Phys. Rev. D, 100, 043027, \dodoi{10.1103/PhysRevD.100.043027}

\bibitem[{Z. Roupas(2025)Roupas}]{Roupas:2025djb}
Roupas, Z. 2025, \bibinfo{title}{{Black hole mass function shift in
  proto-stellar-clusters driven by gas accretion},} Astron. Astrophys., 702,
  A208, \dodoi{10.1051/0004-6361/202556434}

\bibitem[{W.-H. Ruan {et~al.}(2020)Ruan, Guo, Cai, \& Zhang}]{Ruan:2018tsw}
Ruan, W.-H., Guo, Z.-K., Cai, R.-G., \& Zhang, Y.-Z. 2020,
  \bibinfo{title}{{Taiji program: Gravitational-wave sources},} Int. J. Mod.
  Phys. A, 35, 2050075, \dodoi{10.1142/S0217751X2050075X}

\bibitem[{H. Sana {et~al.}(2012)Sana, de~Mink, de~Koter, Langer, Evans, Gieles,
  Gosset, Izzard, Bouquin, \& Schneider}]{Sana:2012px}
Sana, H., de~Mink, S.~E., de~Koter, A., {et~al.} 2012, \bibinfo{title}{{Binary
  interaction dominates the evolution of massive stars},} Science, 337, 444,
  \dodoi{10.1126/science.1223344}

\bibitem[{M.~A. Sedda {et~al.}(2023)Sedda, Naoz, \& Kocsis}]{Sedda:2023big}
Sedda, M.~A., Naoz, S., \& Kocsis, B. 2023, \bibinfo{title}{{Quiescent and
  Active Galactic Nuclei as Factories of Merging Compact Objects in the Era of
  Gravitational Wave Astronomy},} Universe, 9, 138,
  \dodoi{10.3390/universe9030138}

\bibitem[{M.~A. Sedda {et~al.}(2021)Sedda, Rizzuto, Naab, Ostriker, Giersz, \&
  Spurzem}]{Sedda:2021abh}
Sedda, M.~A., Rizzuto, F.~P., Naab, T., {et~al.} 2021,
  \bibinfo{title}{{Breaching the Limit: Formation of GW190521-like and IMBH
  Mergers in Young Massive Clusters},} Astrophys. J., 920, 128,
  \dodoi{10.3847/1538-4357/ac1419}

\bibitem[{M. Spera {et~al.}(2019)Spera, Mapelli, Giacobbo, Trani, Bressan, \&
  Costa}]{Spera:2018wnw}
Spera, M., Mapelli, M., Giacobbo, N., {et~al.} 2019, \bibinfo{title}{{Merging
  black hole binaries with the SEVN code},} Mon. Not. Roy. Astron. Soc., 485,
  889, \dodoi{10.1093/mnras/stz359}

\bibitem[{A. {Stacy} {et~al.}(2016){Stacy}, {Bromm}, \&
  {Lee}}]{2016MNRAS.462.1307S}
{Stacy}, A., {Bromm}, V., \& {Lee}, A.~T. 2016, \bibinfo{title}{{Building up
  the Population III initial mass function from cosmological initial
  conditions},} \mnras, 462, 1307, \dodoi{10.1093/mnras/stw1728}

\bibitem[{J. Stegmann {et~al.}(2025)Stegmann, Olejak, \&
  de~Mink}]{Stegmann:2025cja}
Stegmann, J., Olejak, A., \& de~Mink, S.~E. 2025, \bibinfo{title}{{Resolving
  Black Hole Family Issues among the Massive Ancestors of Very High-spin
  Gravitational-wave Events like GW231123},} Astrophys. J. Lett., 992, L26,
  \dodoi{10.3847/2041-8213/ae0e5f}

\bibitem[{N.~C. Stone {et~al.}(2017)Stone, Metzger, \& Haiman}]{Stone:2016wzz}
Stone, N.~C., Metzger, B.~D., \& Haiman, Z. 2017, \bibinfo{title}{{Assisted
  inspirals of stellar mass black holes embedded in AGN discs: solving the
  {\textquoteleft}final au problem{\textquoteright}},} Mon. Not. Roy. Astron.
  Soc., 464, 946, \dodoi{10.1093/mnras/stw2260}

\bibitem[{H. Tagawa {et~al.}(2020)Tagawa, Haiman, \& Kocsis}]{Tagawa:2019osr}
Tagawa, H., Haiman, Z., \& Kocsis, B. 2020, \bibinfo{title}{{Formation and
  Evolution of Compact Object Binaries in AGN Disks},} Astrophys. J., 898, 25,
  \dodoi{10.3847/1538-4357/ab9b8c}

\bibitem[{A. Tanikawa(2024)Tanikawa}]{Tanikawa:2024mpj}
Tanikawa, A. 2024, \bibinfo{title}{{Contribution of population III stars to
  merging binary black holes},} Rev. Mod. Plasma Phys., 8, 13,
  \dodoi{10.1007/s41614-024-00153-8}

\bibitem[{A. Tanikawa {et~al.}(2025)Tanikawa, Liu, Wu, Fujii, \&
  Wang}]{Tanikawa:2025fxw}
Tanikawa, A., Liu, S., Wu, W., Fujii, M.~S., \& Wang, L. 2025,
  \bibinfo{title}{{GW231123 Formation from Population III Stars: Isolated
  Binary Evolution},} \doarXiv{2508.01135}

\bibitem[{A. Tanikawa {et~al.}(2020)Tanikawa, Yoshida, Kinugawa, Takahashi, \&
  Umeda}]{Tanikawa:2019isi}
Tanikawa, A., Yoshida, T., Kinugawa, T., Takahashi, K., \& Umeda, H. 2020,
  \bibinfo{title}{{Fitting formulae for evolution tracks of massive stars under
  extreme metal-poor environments for population synthesis calculations and
  star cluster simulations},} Mon. Not. Roy. Astron. Soc., 495, 4170,
  \dodoi{10.1093/mnras/staa1417}

\bibitem[{A. Tanikawa {et~al.}(2022)Tanikawa, Yoshida, Kinugawa, Trani,
  Hosokawa, Susa, \& Omukai}]{Tanikawa:2021qqi}
Tanikawa, A., Yoshida, T., Kinugawa, T., {et~al.} 2022, \bibinfo{title}{{Merger
  Rate Density of Binary Black Holes through Isolated Population I, II, III and
  Extremely Metal-poor Binary Star Evolution},} Astrophys. J., 926, 83,
  \dodoi{10.3847/1538-4357/ac4247}

\bibitem[{M.~P. Vaccaro {et~al.}(2024)Vaccaro, Mapelli, P{\'e}rigois, Barone,
  Artale, Dall'Amico, Iorio, \& Torniamenti}]{Vaccaro:2023cwr}
Vaccaro, M.~P., Mapelli, M., P{\'e}rigois, C., {et~al.} 2024,
  \bibinfo{title}{{Impact of gas hardening on the population properties of
  hierarchical black hole mergers in active galactic nucleus disks},} Astron.
  Astrophys., 685, A51, \dodoi{10.1051/0004-6361/202348509}

\bibitem[{L. {Wang} {et~al.}(2020){Wang}, {Iwasawa}, {Nitadori}, \&
  {Makino}}]{2020MNRAS.497..536W}
{Wang}, L., {Iwasawa}, M., {Nitadori}, K., \& {Makino}, J. 2020,
  \bibinfo{title}{{PETAR: a high-performance N-body code for modelling massive
  collisional stellar systems},} \mnras, 497, 536,
  \dodoi{10.1093/mnras/staa1915}

\bibitem[{L. Wang {et~al.}(2022)Wang, Tanikawa, \& Fujii}]{Wang:2022unj}
Wang, L., Tanikawa, A., \& Fujii, M. 2022, \bibinfo{title}{{Gravitational wave
  of intermediate-mass black holes in Population III star clusters},} Mon. Not.
  Roy. Astron. Soc., 515, 5106, \dodoi{10.1093/mnras/stac2043}

\bibitem[{S.~E. Woosley \& A. Heger(2021)Woosley \& Heger}]{Woosley:2021xba}
Woosley, S.~E., \& Heger, A. 2021, \bibinfo{title}{{The Pair-Instability Mass
  Gap for Black Holes},} Astrophys. J. Lett., 912, L31,
  \dodoi{10.3847/2041-8213/abf2c4}

\bibitem[{W. {Wu} {et~al.}(2025){Wu}, {Wang}, {Liu}, {Sun}, {Tanikawa}, \&
  {Fujii}}]{2025ApJ...986..163W}
{Wu}, W., {Wang}, L., {Liu}, S., {et~al.} 2025,
  \bibinfo{title}{{Pair-instability Gap Black Holes in Population III Star
  Clusters: Pathways, Dynamics, and Gravitational-wave Implications},} \apj,
  986, 163, \dodoi{10.3847/1538-4357/add1df}

\bibitem[{Y. Yang {et~al.}(2019)Yang {et~al.}}]{Yang:2019cbr}
Yang, Y., {et~al.} 2019, \bibinfo{title}{{Hierarchical Black Hole Mergers in
  Active Galactic Nuclei},} Phys. Rev. Lett., 123, 181101,
  \dodoi{10.1103/PhysRevLett.123.181101}

\bibitem[{C. Yuan {et~al.}(2025)Yuan, Chen, \& Liu}]{Yuan:2025avq}
Yuan, C., Chen, Z.-C., \& Liu, L. 2025, \bibinfo{title}{{GW231123 mass gap
  event and the primordial black hole scenario},} Phys. Rev. D, 112, L081306,
  \dodoi{10.1103/2vfn-48kh}

\end{thebibliography}
\bibliographystyle{aasjournalv7}


\end{CJK*}
\end{document}